\documentclass[12pt]{article}
\usepackage{mymacros}

\title{Axion-de Sitter wormholes }
\author[a]{Sergio E. Aguilar-Gutierrez,}
\author[a]{Thomas Hertog,}
\author[a]{Rob Tielemans,}
\author[b,c]{\\ Jan Pieter van der Schaar,}
\author[a]{Thomas Van Riet}
\affiliation[a]{Institute for Theoretical Physics, KU Leuven, Celestijnenlaan 200D, B-3001 Leuven, Belgium}
\affiliation[b]{Institute of Physics, University of Amsterdam, Science Park 904, PO Box 94485, 1090 GL Amsterdam, The Netherlands}
\affiliation[c]{Delta Institute for Theoretical Physics, Science Park 904, PO Box 94485, 1090 GL Amsterdam, The Netherlands}
\emailAdd{sergio.ernesto.aguilar@gmail.com, thomas.hertog@kuleuven.be, rob.tielemans@kuleuven.be, j.p.vanderschaar@uva.nl, thomas.vanriet@kuleuven.be}
\abstract{We construct wormholes supported by axion flux in the presence of a positive cosmological constant. The solutions describe compact, one-handle bodies colloquially known as kettlebell geometries. The wormholes are perturbatively stable, but regularity of the Euclidean geometry implies an upper bound on the axion flux. Viewed as no-boundary saddle points, wormholes are suppressed relative to the round sphere. The symmetric kettlebell with maximal axion density has vanishing Euclidean action. Continuing into the Lorentzian across the equator, the solutions describe two expanding branches of de Sitter space filled with an axion field that rapidly dilutes and which are connected by a quantum bounce across which the arrow of time reverses. }

\begin{document}

\maketitle

\section{Introduction}\label{sec:intro}
Does a consistent theory of quantum gravity allow for topology change? During such transitions, a universe with a given spatial geometry evolves in a non-diffeomorphic way by some quantum mechanical process. The transition amplitude for such a process can in principle be computed by using a saddle point expansion of the path integral of Euclidean quantum gravity. However, quantum transitions between manifolds with different topology appears to pose several paradoxes: they could break locality, lead to the loss of quantum coherence with respect to a given observer, or alternatively, these processes might give rise to intrinsic randomness of the coupling constants in nature \cite{Coleman:1988cy, Lavrelashvili:1987jg, Hebecker:2018ofv}. That is why it has been suggested that these effects should be absent in any consistent theory of quantum gravity and indeed there are some indications that such topology-changing processes are absent in string theory \cite{McNamara:2019rup}.

The aim of this work is to study axion wormholes in a universe with a positive cosmological constant\footnote{So we will not be able to say anything about the UV completion, which has been explored for flat and AdS wormholes \cite{Bergshoeff:2004pg, Hertog:2017owm, Katmadas:2018ksp,Loges:2023ypl}. UV completion requires at least adding extra scalars, like saxions which are typically massive in a phenomenological context \cite{Andriolo:2022rxc, Jonas:2023ipa} and can be massive or massless in holographic contexts \cite{Hertog:2017owm, Katmadas:2018ksp, Loges:2023ypl}. }. 
The important details and corresponding interpretation of this particular case have apparently been largely overlooked since the work of \cite{Gutperle:2002km}, where they constructed the general background solution. 

Our starting point is Einstein gravity with a positive cosmological constant in the presence of axion fluxes, in $3$ or more dimensions. We analyze the topology of the smooth Euclidean background solutions, and conclude that the effect of the axion flux is to nucleate a handle out of the Euclidean sphere, resulting in a `kettlebell'. In addition, the space of allowed (real Euclidean) solutions reveals a bound on the maximal axion flux threading the co-dimension $1$ sphere, depending on the de Sitter length scale. In analogy with Schwarzschild-de Sitter black hole solutions, we call the largest, critical, solutions ``Nariai" wormholes, for which the size of the fibered sphere is constant along the Euclidean time circle.  

We then compute the on-shell Euclidean action of these axionic de Sitter wormholes. Notably, the on-shell action is well-approximated in terms of a linear function of the axion flux parameter, in arbitrary dimensions. A feature again analogous to the (approximate) linear mass dependence of the on-shell action for Schwarzschild-de Sitter black holes \cite{Morvan:2022ybp}. Under closer scrutiny our results also reveal that a partitioning of the axion flux does not lower the action, preventing fragmentation into multiple smaller axion wormholes. We also confirm that, for fixed axion flux, these solutions are perturbatively stable, and consistently reproduce the flat space axion wormhole results in the limit of a vanishing cosmological constant \cite{Loges:2022nuw}\footnote{The literature on wormhole stability contains contradictory statements in the case of a non-positive cosmological constant \cite{Hertog:2018kbz}, mostly because a reliable gauge invariant analysis with proper boundary conditions was lacking until the substantial improvement in \cite{Loges:2022nuw,progress}.}.

We then study Lorentzian continuations \cite{Hartle:1983ai} of the wormhole solutions. A priori there are several possibilities to cut and glue these Euclidean solutions to Lorentzian ones. In the Lorentzian regime, the presence of axion flux effectively acts as a cosmological fluid with an ultra-stiff equation of state $w=1$. Therefore for large scale factors, the Lorentzian cosmologies reduce to pure de Sitter. Notably, however, in the presence of axion flux a new (singular) regime of solutions opens up for very small scale factors. This regime is separated from the large scale factor solution by a barrier in the effective potential\footnote{This was also noticed in \cite{Bouhmadi-Lopez:2017sgq,Chen:2016ask} where a link with wormholes was suggested.}. Despite these complications, we will advance an interpretation of these wormholes as saddles of a variant of the Hartle-Hawking no-boundary wave function. In this interpretation, the wormholes can be viewed as quantum bridges connecting two expanding branches of global de Sitter-like universes, across which the arrow of time reverses.

\section{de Sitter wormholes}\label{sec:Wormhole solutions}

We consider gravity coupled to an axion field $\chi$ in $d>2$ dimensions governed by the Euclidean action
\begin{equation}\label{eq:onshell fundamental}
I[g, B_{d-2}]=\int\qty[-\tfrac{1}{2\kappa_d^2}\star(R-2\Lambda)+\tfrac{1}{2}\star H_{d-1}\wedge H_{d-1}]~,
\end{equation}
where $H_{d-1}=dB_{d-2}$, $\kappa_d^2=8\pi G_N$ and the cosmological constant
\begin{equation}\label{eq:Lambda C.C.}
    \Lambda=\tfrac{(d-1)(d-2)}{2\ell^2}>0~.
\end{equation}
To Hodge dualize the gauge potential $B_{d-2}$ we regard the action as a functional of $H$ instead and add a Lagrange multiplier $\chi$ enforcing the closure of $H$ via the extra term $\chi\rmd H_{d-1}$. Then we find  $H=\star \rmd\chi$.
In terms of the axion dual $\chi$, the action (\ref{eq:onshell fundamental}) reads
\begin{equation}\label{eq:onshell axionic}
    I[g, \chi]=\int\qty[-\tfrac{1}{2\kappa_d^2}\star(R-2\Lambda)-\tfrac{1}{2}\star\rmd\chi\wedge\rmd\chi-\rmd(\chi\star\rmd\chi)].
\end{equation}
Despite appearances the actions (\ref{eq:onshell fundamental}) and (\ref{eq:onshell axionic}) are equivalent on–shell. This is because even though (\ref{eq:onshell fundamental}) has a manifestly positive kinetic energy for $H_{d-1}$ whereas (\ref{eq:onshell axionic}) contains a negative kinetic energy term for $\chi$, the action (\ref{eq:onshell axionic}) also contains a total derivative term. For a given background manifold that supports axion flux, the latter acquires a monodromy upon integration. This yields the positive contribution of $H_{d-1}$ to the Euclidean energy in the formalism using the axion scalar $\chi$.

\subsection{Euclidean geometry}\label{eq:geometry}

The metric of Euclidean, spherically symmetric axion wormholes in $d$ dimensions can be written in the following form \cite{Gutperle:2002km},
\begin{align} \label{scale}
    &\rmd s^2=N(\tau)^2\rmd\tau^2+a(\tau)^2\rmd\Omega_{d-1}^2\,.
\end{align}
where the Euclidean time $\tau$ lives in a finite interval $\tau\in[\tau_{\rm min},\,\tau_{\rm max}]$, the endpoints of which are identified. Axion flux consistent with the symmetry comes in the form
\begin{equation}
  H_{d-1}=Q\,
  \text{Vol}(\text{S}^{d-1})~,
  \label{eq:axion solutions}
\end{equation}
where the constant $Q$ denotes the flux density and $\text{Vol}(\text{S}^{d-1})$ is the $(d-1)-$form volume element of the sphere $\rmd s^2_{(d-1)}=\rmd\Omega_{d-1}^2$.

The Einstein equations imply that the Euclidean scale factor $a(\tau)$ obeys the following constraint,
\begin{equation}
    \frac{1}{N} \frac{\rmd a}{\rmd \tau} =\pm \sqrt{1-\frac{a^2}{\ell^2} -\frac{\kappa_d^2 Q^2 \,a^{-2(d-2)}}{(d-1)(d-2)}}\label{eq:derivative r, tau}\,.
\end{equation}
which, for a given gauge choice, yields an explicit expression for the metric. 

For $Q=0$, the solution is simply the $d$-sphere, viz. Euclidean de Sitter space. For $Q>0$, the above constraint implies that the scale factor oscillates between a positive minimum and maximum value. Upon identification of the endpoints of the interval, this generates a `kettlebell' geometry that is topologically $S^1 \times S^{d-1}$, viz. a Euclidean wormhole. The minimum value $a_{\rm min}$ of the scale factor sets the size of what we will call the wormhole throat, while the maximum value $a_{\rm max}$ corresponds to the equator of the parent Euclidean de Sitter space. For increasing $Q$ the size of the wormhole throat approaches that of the dS equator. For the maximally allowed axion charge the scale factor in \eqref{scale} becomes everywhere constant and the solution reduces to,
\begin{equation}\label{torus} 
    \rmd s^2=\rmd \tau^2+\frac{d-2}{d-1}\ell^2\rmd\Omega^2_{d-1}\,. 
\end{equation}
In a way that is reminiscent of the Einstein static universe, the axion density and the constant value of the scale factor are both determined by the value of the cosmological constant:
\begin{align}
    Q^2_{\text{max}}&=\frac{\ell^{2(d-2)}}{\kappa^2_d}(d-2)\qty(\frac{d-2}{d-1})^{d-2}\,,\label{eq:max size WH}\\
    a_{\text{max}}(Q_\text{max})&=a_{\text{min}}(Q_\text{max})=\sqrt{\frac{d-2}{d-1}}\ell\,,\label{eq:Conf scaling factor Nariai}
\end{align}
We will refer to this limiting solution as the 'Nariai' wormhole because it is somewhat analogous to the largest, Nariai black hole that fits in de Sitter space.

\begin{figure}[t]
    \centering
     \subfloat[]{\includegraphics[width=0.5\textwidth]{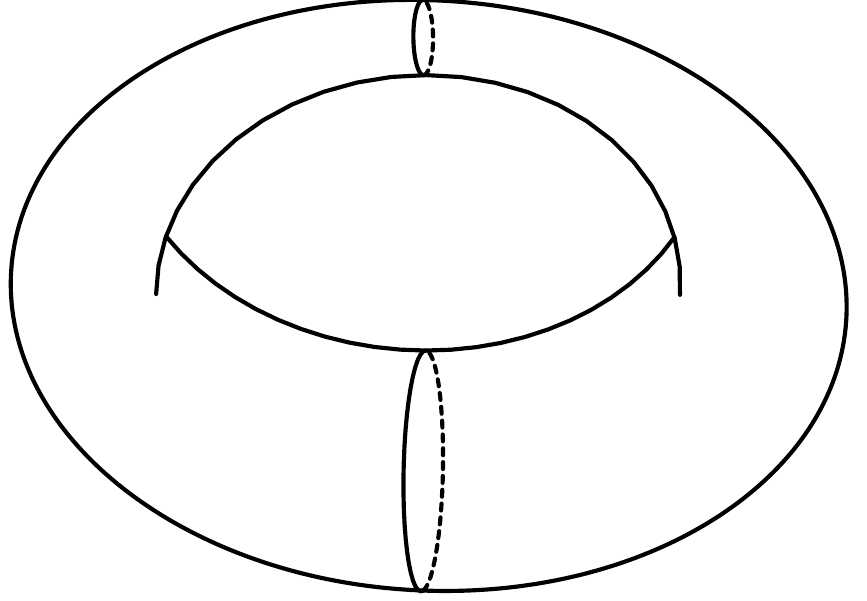}}\\
    \subfloat[]{\includegraphics[width=0.6\textwidth]{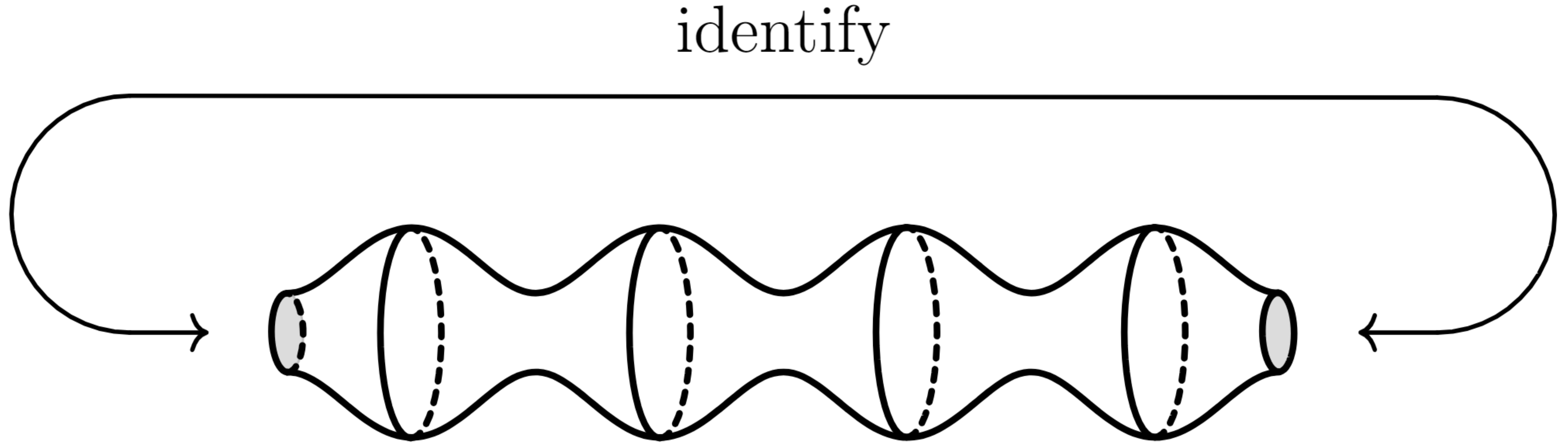}}
    \caption{(a) The Euclidean wormhole has an S$^1\times$S$^{d-1}$ topology, and is known as a kettlebell geometry. (b) Euclidean beads joined by multiple wormhole throats.}
    \label{fig:handles}
\end{figure}

Finally, we note that the wormhole metric can be obtained explicitly in $d=3$ dimensions. It is given by 
\begin{equation}\label{threed}
    \rmd s^2=\rmd \tau^2+\frac{\ell^2}{2}\qty(1 + \sin(2\frac{\tau}{\ell})\sqrt{1 -\frac{2\kappa_3^2 Q^2}{\ell^2}})\rmd\Omega_2^2~.
\end{equation}
Hence
\begin{equation}
a^2_{\rm max, \rm min}=\frac{\ell^2}{2}\qty(1\pm\sqrt{1 - \frac{2\kappa_3^2 Q^2}{\ell^2}} ),    
\end{equation}
with 
\begin{equation}
2\tau/\ell\in\qty[-\frac{\pi}{2},\,\frac{3\pi}{2}], \label{eq:allowed range d3}   
\end{equation}
and with the endpoints of this interval identified. For $Q=0$ we have $a_{\rm min}=0$, so the endpoints turn into the poles of the three-sphere. Note also that for $Q>0$ it is straightforward to construct a necklace of wormholes by extending the range of $\tau$ to cover more than one complete oscillation of the scale factor. We depict an example in figure \ref{fig:handles}.
Such necklaces of wormholes are of course possible in arbitrary dimensions $d$.

\subsection{On-shell action}
The on-shell Euclidean action $I_E$ of wormholes of the form \eqref{scale} is given by 
\begin{equation}\label{eq:action to be gauged}
I_{\rm E}[Q]={\Omega_{d-1}}\int \rmd\tau\,N(\tau)a(\tau)^{d-1}\qty(\frac{1-d}{\kappa_d^2\ell^2}+\frac{Q^2}{2a(\tau)^{2(d-1)}})~.
\end{equation}
Substituting (\ref{eq:derivative r, tau}) in (\ref{eq:action to be gauged}) yields
\begin{equation}
I_{\rm E}[Q]=2{\Omega_{d-1}}\int_{a_{\text{min}}}^{a_{\text{max}}}\rmd a\,a^{d-1} \frac{\qty(\frac{1-d}{\kappa_d^2\ell^2}+\frac{Q^2}{2a^{2(d-1)}})}{\sqrt{1-\frac{a^2}{\ell^2}-\frac{\kappa_d^2Q^2}{(d-1)(d-2)\,a^{2(d-2)}}}}\,. \label{eq:vol}
\end{equation}

Evaluating this integral in $d=3$, using the explicit solution \eqref{threed}, gives the following closed-form expression, 
\begin{align}
    \eval{I_{\rm E}[Q]}_{d=3}=-\frac{4\pi^2\ell}{\kappa_3^2}\qty(1-\frac{\sqrt{2}\kappa_3Q}{\ell})~.\label{eq:onshell d=3}
\end{align}
We see that the on-shell Euclidean action is always negative and that it increases linearly in $Q$. Viewed as saddle points, therefore, the wormholes are suppressed relative to the three-sphere. Further, the action of the limiting `Nariai wormhole' vanishes. Also, the flat limit of (\ref{eq:vol}) can be evaluated and reproduces the Giddings-Strominger expression \cite{Giddings:1989bq}.
\begin{figure}[t!]
    \centering
    \subfloat[]{\includegraphics[width=0.49\textwidth]{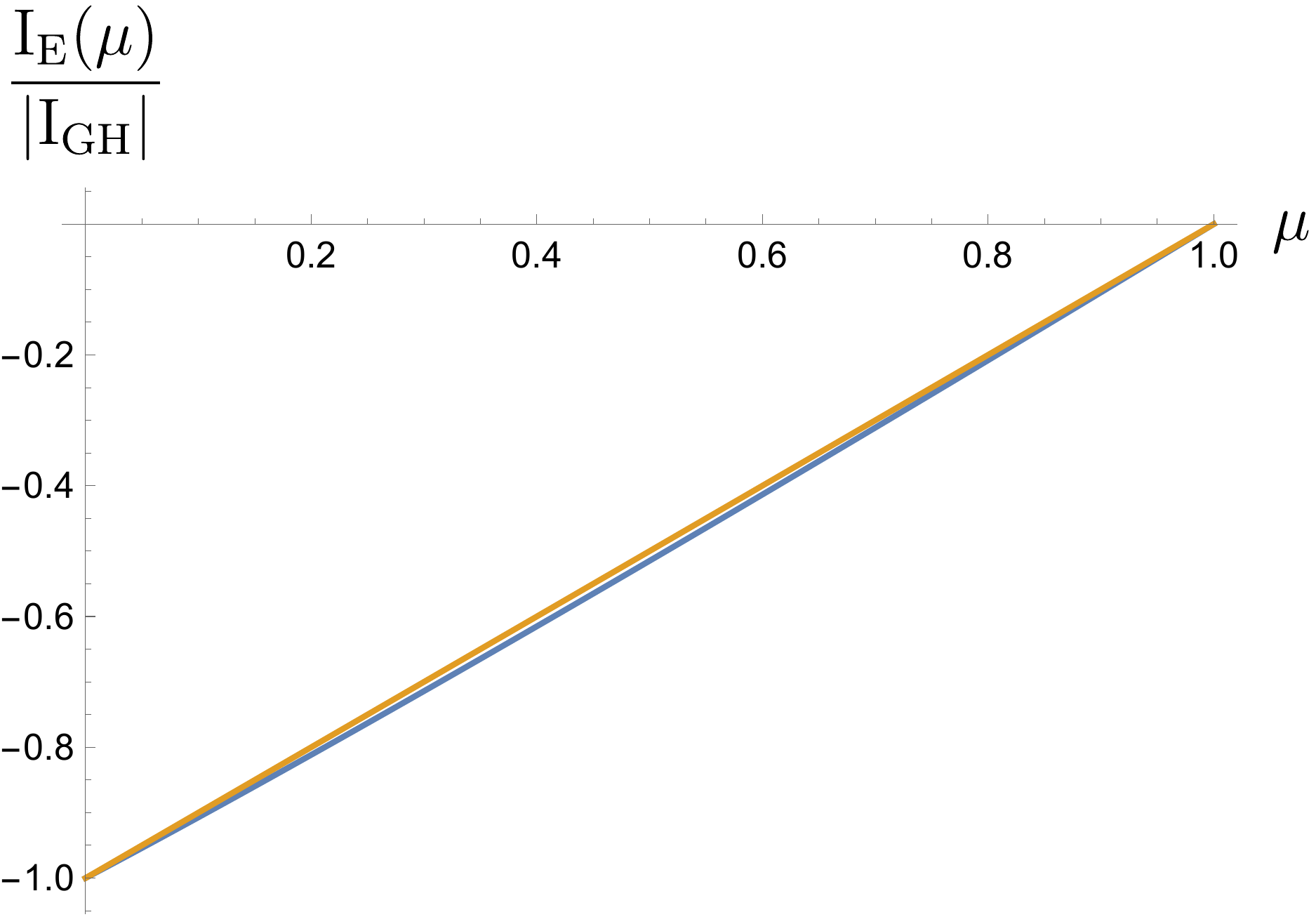}}\subfloat[]{\includegraphics[width=0.49\textwidth]{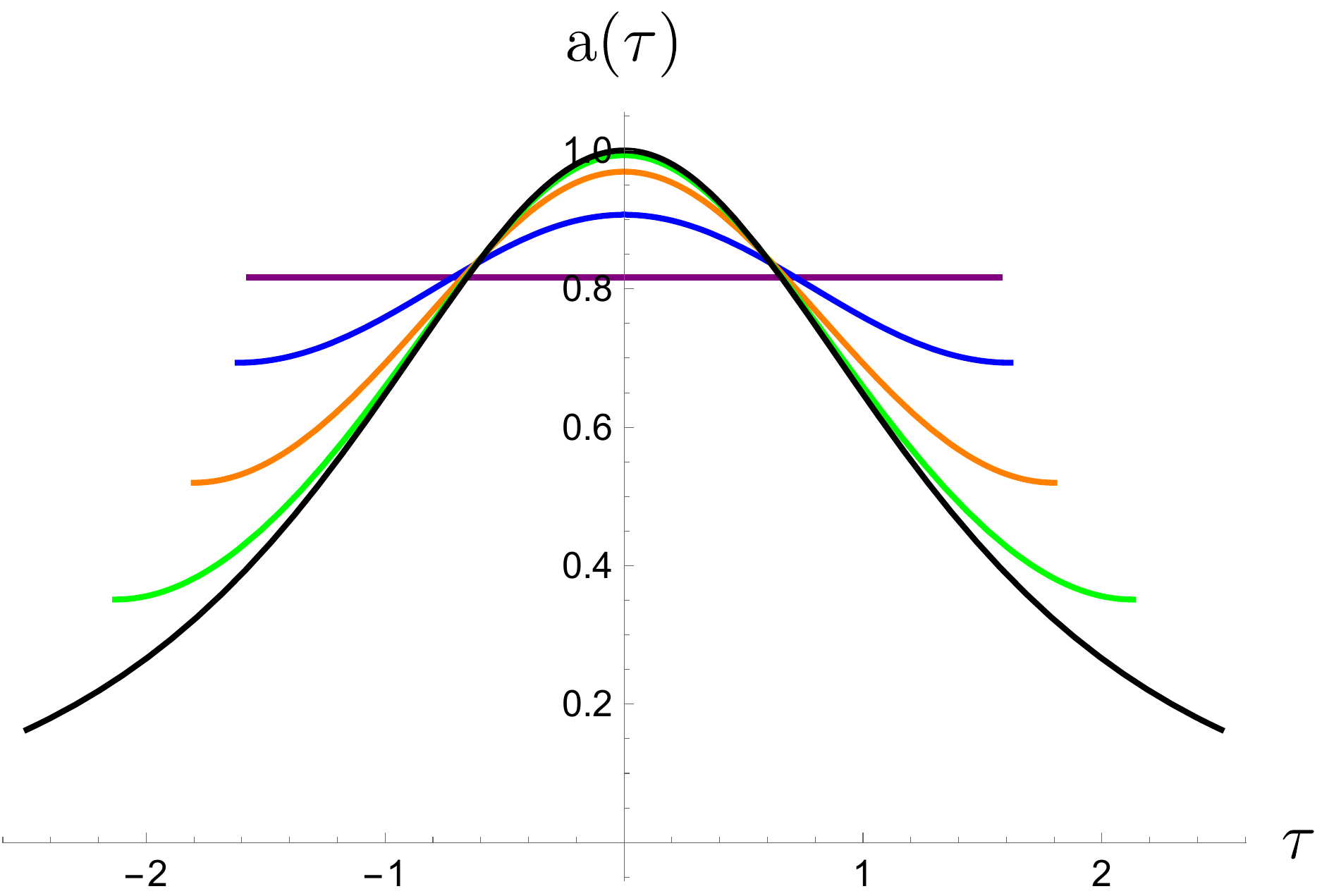}}
    \caption{(a) The on-shell action of generic Euclidean dS wormholes in $d=4$, normalized by the absolute value of the Gibbons-Hawking action (\ref{eq:GH in d dim}) of the four-sphere. The action is shown here in terms of the parameter $\mu\equiv Q/Q_{\rm max}$. The blue curve is the exact numerical integration of (\ref{eq:vol}); the orange curve employs the linear approximation (\ref{eq:nice OnShell action}). We see that the action of the Nariai wormhole vanishes exactly. (b) The scale factor in conformal gauge $N(\tau)=a(\tau)$ in (\ref{scale}), for different values of the axion charge ratio, viz. $\mu=1$ (purple), $0.9$ (blue), $0.6$ (orange), $0.3$ (green), and finally $\mu=0$ (black), corresponding to the four-sphere. We refer to the location of the minimum of the scale factor as the wormhole throat, chosen here at $\tau=\tau_{\rm min}$, and $\tau=\tau_{\rm max}=-\tau_{\rm min}$.}
    \label{fig:onshell full}
\end{figure}

Next, we evaluate the integral (\ref{eq:vol}) numerically in $d=4$. Figure \ref{fig:onshell full} shows the result as a function of the flux parameter $Q$. One sees that $I_{\rm E}[Q]_{d=4}$ too behaves almost exactly linearly, and that the action of the four-dimensional Nariai wormhole vanishes. It turns out these are properties of the on-shell action of wormholes in general dimensions. In effect, an accurate approximate expression of $I_{\rm E}[Q]_{d}$ in any number of dimensions $d$ is given by
\begin{equation}\label{eq:nice OnShell action}
    I_{\rm E}[\mu]\approx I_{\rm GH}\qty(1-\mu),
\end{equation}
where $\mu\equiv Q/Q_{\rm max}$, with
\begin{align}
Q^2_{\text{max}}&=\frac{\ell^{2(d-2)}}{\kappa^2_d}(d-2)\qty(\frac{d-2}{d-1})^{d-2}\,,\\
    I_{\rm GH}&=-\frac{4\pi^{(d+1)/2}\ell^{d-2}}{\kappa^2_d\,\Gamma\qty(\tfrac{d-1}{2})}\,.\label{eq:GH in d dim}
\end{align}
Finally we note that $\dv[2]{I_{\rm E}(Q)}{Q}>0$ in $d=4$, as illustrated in Fig. \ref{fig:onshell full}. This implies that the single wormholes are non-perturbatively stable against fragmentation into a necklace of $n$ wormholes with $n \, q=Q$. Generalizing the results of \cite{Loges:2022nuw, progress}, we verified in Appendix \ref{sec:Perturbative Stability} that the wormhole solutions, for fixed $Q$, are also perturbatively stable. As a consequence, it is, therefore, reasonable to assume they represent physical saddles. We consider this in the next section. 

\section{Axionic quantum bounces}\label{sec:Interpretation}

Euclidean axion wormholes are so-called real tunneling geometries. That is, they possess a hypersurface of vanishing extrinsic curvature on which they can be glued onto a Lorentzian solution. In effect, axion-de Sitter wormholes have two such hypersurfaces, as indicated in Figure \ref{fig:handles}, namely the wormhole throat and the equator of the parent sphere. The Lorentzian evolution(s) that emerge from the initial conditions specified by the Euclidean geometry are governed by the Wick-rotated Friedmann-Lema\^itre equation \eqref{eq:derivative r, tau}. Working in conformal gauge (i.e. $N(t)=a(t)$) this can be written as
\begin{align}\label{FL}
    \left(\frac{\rmd a}{\rmd t}\right)^2 + V_{\rm eff}(a) = 0,
\end{align}
where the effective potential 
\begin{align}\label{eff}
V_{\rm eff}(a) = a^2-\frac{a^4}{\ell^2}-\frac{\kappa_d^2Q^2a^{-2(d-3)}}{(d-1)(d-2)}
\end{align}

\begin{figure}
    \centering
\includegraphics[width=0.7\textwidth]{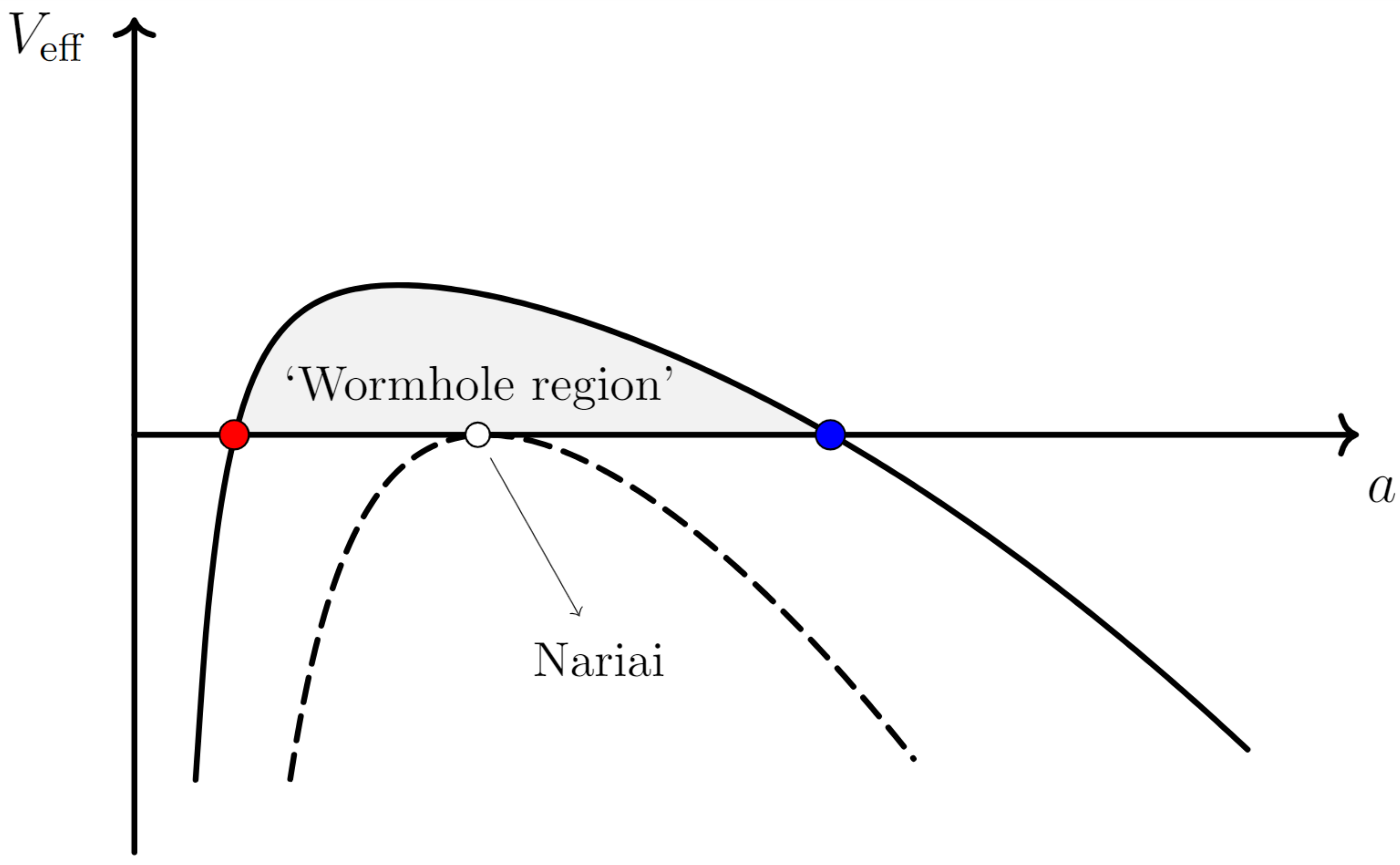}
    \caption{Effective potential for the scale factor in conformal gauge. For subcritical values of $Q$ (solid line), Euclidean wormholes correspond to solutions in the region under the barrier as indicated. The turning points correspond to the wormhole throat at $a=a_{\rm min}$ (red) and to the equator of the parent sphere at $a=a_{\rm max}$ (blue), as in Figure \ref{fig:handles}. The Nariai wormhole arises for a critical value of $Q$ (dashed line) and describes an Einstein static universe in the Lorentzian. Lorentzian cosmologies traverse scale factor values outside the barrier region. To the right, there is a history that bounces at a minimum radius, whereas to the left there is one bouncing at a maximum size.}
    \label{fig: effecive potential}
\end{figure}
 This Lorentzian system \eqref{eff} effectively consists of gravity coupled to a positive cosmological constant and an ultra-stiff fluid with an equation of state parameter $w=1$ that arises from the $Q$ units of axion flux through the $(d-1)$-sphere. The effective potential is shown in Figure \ref{fig: effecive potential}. We can read off the two homogeneous and isotropic cosmologies that the theory admits for a given value of $Q$ and where they can be patched onto the Euclidean geometry that captures the quantum regime under the potential barrier. One solution bounces at a minimum radius and is approximately de Sitter space. However, the axion flux introduces a second classical turning point, at a value of the scale factor where it equals the size of the wormhole throat. This gives rise to a Lorentzian cosmology with an over-critical axion density, which bounces when it reaches a maximum radius before collapsing into a big crunch singularity. In the special case of maximum axion flux, both turning points coincide and hence the region under the barrier vanishes. This corresponds to the Nariai wormhole in the Euclidean, which, as we have seen, has vanishing on-shell action. The corresponding Lorentzian solution at the maximum of the effective potential is an (unstable) static universe. 
\begin{figure}
\subfloat[ ]{\includegraphics[width=0.25\textwidth]{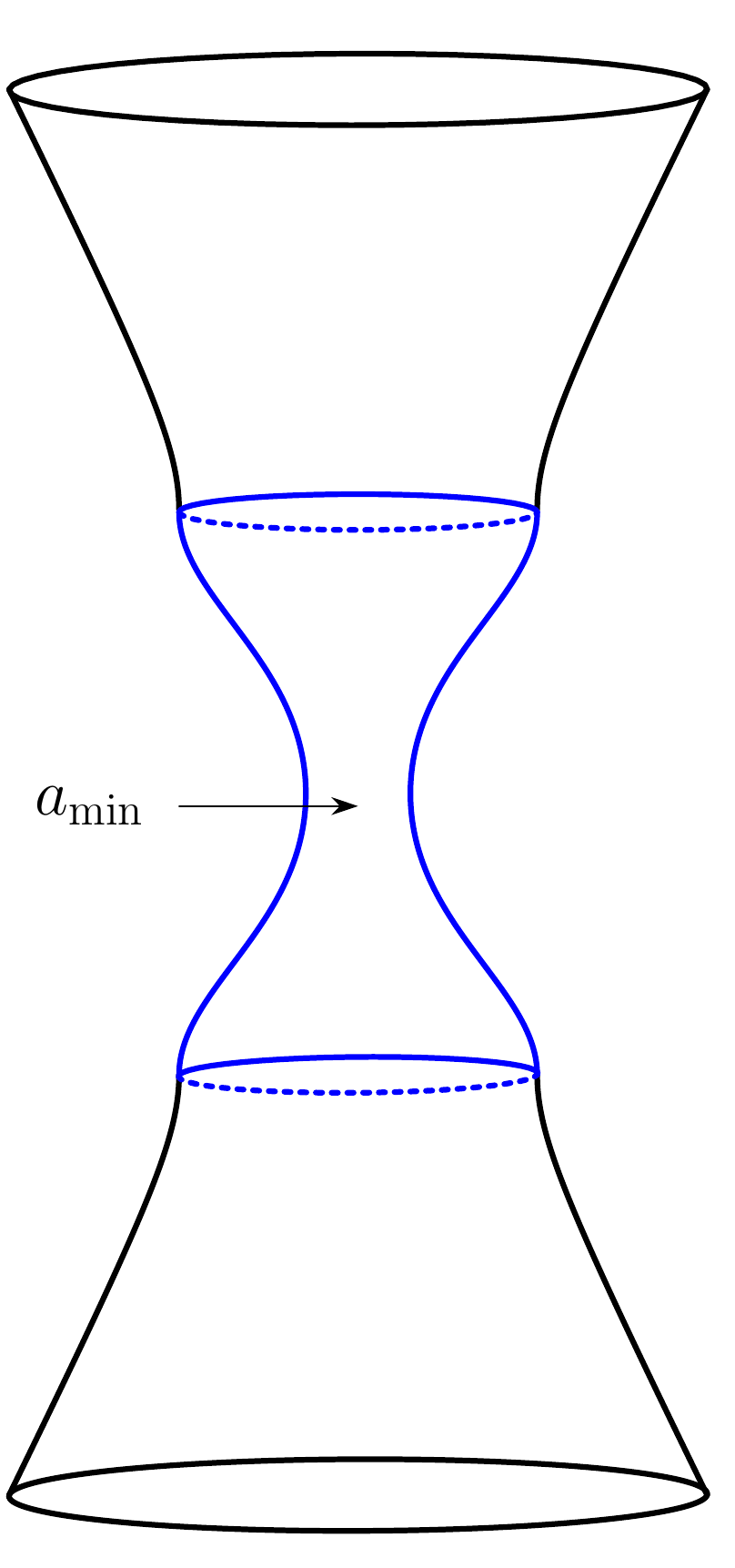}}\hspace{0.5cm} \subfloat[ ]{\includegraphics[width=0.3\textwidth]{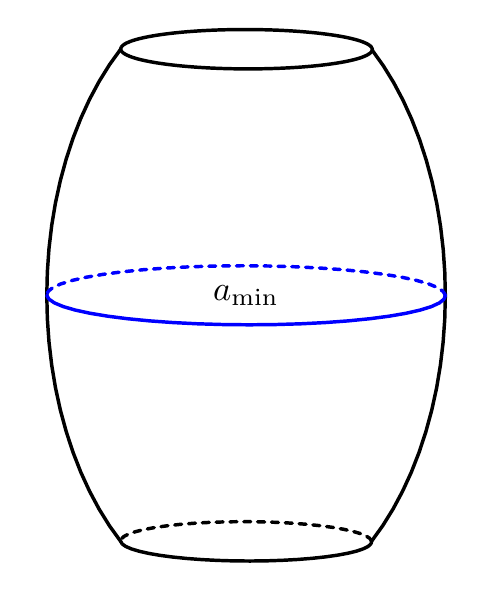}}
\caption{The homogeneous and isotropic minisuperspace, for fixed axion charge and in the presence of a positive cosmological constant, encompasses two qualitatively different cosmological histories. Panel (a) shows an asymptotically de Sitter universe associated with a wormhole saddle that mediates a quantum bounce. Panel (b) shows a history that emerges from a big bang and recollapses into a big crunch, bouncing when its size reaches that of the wormhole throat. The no-boundary state favors the former over the latter in this model, i.e. the wave function decays towards small values of the scale factor. Note that the scale factor in panel (b) is rescaled relative to the one in (a).}
\label{subfig:HHA-instanton}
\end{figure}

The Euclidean wormhole can be thought of as preparing the state of these Lorentzian cosmologies. More precisely, the wormholes provide a saddle point approximation to a quantum state of the universe that includes these cosmological histories. Since the Euclidean geometries are regular and compact, without boundary, it is natural to regard the wormholes as valid saddle points of the Hartle-Hawking no-boundary wave function in a two-dimensional minisuperspace consisting of two $(d-1)$-spheres of size $a_1$, $a_2$ with $Q_1$ and $Q_2$ units of axion flux going through respectively. In this interpretation, a given wormhole specifies the amplitude of a pair of expanding axion-de Sitter universes. 

In the familiar context of scalar field inflation, no-boundary saddles come in complex conjugate pairs. Both members of every pair specify the amplitude of the expanding branch of the same inflationary universe. Even though classically such inflationary universes are connected across a de Sitter-like bounce at a minimum radius, the complex conjugate saddles that provide the amplitude of each expanding branch are separate geometries. From an Euclidean viewpoint, this is because turning on the inflaton field deforms the sphere, but it obviously doesn't create a handle. The wormhole saddle points differ in this respect in that they provide a quantum mechanical bridge that connects both expanding branches of the approximate de Sitter histories that feature in this model. In the limit of zero flux parameter $Q$, this instanton should reduce to (two copies of) the Hartle-Hawking no-boundary solution. 

This being said, from a physical point of view the saddle point histories with axion flux bear a strong resemblance to the inflationary histories. A key property of the inflationary cosmologies predicted by the no-boundary wave function is that the arrow of time defined by the growth in fluctuations is bidirectional, pointing away from the bounce in both directions \cite{Hartle:2011rb}. This feature ultimately stems from the regularity condition at the South Pole of the saddles that implies fluctuation must vanish there. In Appendix \ref{sec:Arrow of time} we show that, likewise, the fluctuation arrow of time in the axion-de Sitter universes that we consider here, points away from the wormhole in both directions. 

Further evidence for the interpretation of these wormholes as saddles of the no-boundary wave function in an axionic minisuperspace model comes from the behavior of the wave function in the regime under the barrier, for values $a$ of the scale factor in the range $a_{\rm min} < a< a_{\rm max}$. 
In this range, the wormhole throat is retained, but the Euclidean geometry is cut off when the scale factor reaches its boundary value. Consequently, one expects the wave function to decrease for decreasing values of $a$ towards $a_{\rm min}$. This behavior in the Euclidean is the hallmark of the Hartle-Hawking wave function \cite{Hartle:1983ai}. The main difference with the Hartle-Hawking behavior in the context of gravity coupled to a scalar field is that the zero scale factor is replaced with a positive minimum value $a_{\rm min}$, below which a new Lorentzian regime opens up. A WKB analysis of this system with such boundary condition was in fact performed earlier in \cite{Hertog:2021jyd}, in a different context. The saddle point wave function in this new small scale factor regime remains suppressed and involves purely Lorentzian saddles that bounce at a maximum radius $a=a_{\rm min}$, as Figure \ref{subfig:HHA-instanton} illustrates. In some sense, one could say that the axion flux shifts the Hartle-Hawking condition at zero scale factor to a finite value $a=a_{\rm min}$, without significantly altering the behavior of the wave function in this regime of small scale factor. 

\section{Discussion}\label{sec:Discussion}

We have studied Euclidean axion wormholes in $d$ dimensions in the presence of a positive cosmological constant. In the presence of a zero or negative cosmological constant, wormholes describe a smooth bridge that connects two distinct asymptotically flat or AdS spaces. Axion-de Sitter wormholes are qualitatively different. They rather describe a single spherical body with a handle known as a kettlebell geometry. 

The axion flux $Q$ determines the size of the handle - the wormhole throat - relative to the scale set by the cosmological constant. For $Q=0$ the wormhole bridge disappears and the solution reduces to a sphere. Regularity of the geometry also implies an upper bound on $Q$ for which the size of the wormhole throat equals the diameter of the sphere, a limit which we have dubbed the ``Nariai" wormhole.

We found that axion-de Sitter wormholes are perturbatively stable against fluctuations that preserve the axion flux. This is in line with the results of the stability analysis of wormholes in flat and AdS spaces \cite{Loges:2022nuw, progress}. Our expression for the on-shell action furthermore suggests that the wormholes are stable with respect to fluctuations that can alter the local flux, say by fragmenting the wormholes into several smaller ones. We verified for instance that in $d>3$ such necklaces of wormholes have a higher action for the same amount of total axion flux.

Unlike axion wormholes in flat and AdS space, their dS counterparts admit a rather straightforward interpretation as saddles of a variation of the no-boundary wave function. Consider a two-dimensional minisuperspace parameterized by the sizes $a_1$ and $a_2$ of two $(d-1)$-spheres with $Q_1$ and $Q_2$ units of axion flux going through. In this model, the dS wormholes are saddles specifying the amplitude of a pair of identical, asymptotically de Sitter universes with a certain amount of (diluting) axion flux piercing the spatial dimensions. The wormholes essentially provide a quantum bridge that connects the two expanding branches of these dS-like universes. The key difference with the familiar no-boundary wave function in the context of inflation, besides the bridge, is that there is a tail at very small scale factor values where the wave function seemingly predicts Lorentzian behavior in the form of an over-critical, recollapsing universe. That regime is possibly spurious. It is as if the axion flux blows up the zero scale factor that appears in the original wave function to a small but finite value where the no-boundary condition is then imposed.

With regard to the well-known paradoxes associated with axion wormholes, reviewed in \cite{Hebecker:2018ofv}, the wormholes appear to pose a dS version of the ``factorisation paradox"\footnote{We thank the referee for pointing this out to us.} \footnote{See \cite{Maldacena:2004rf} for the original formulation of this problem.} As mentioned, the wormholes may specify the saddle point approximation of a generalization of the Hartle-Hawking no-boundary wave function defined on two $(d-1)$-spheres. In this context, they describe two Lorentzian de Sitter universes with axion charges $Q$ and $-Q$ and connected by a quantum bridge. In the limit of vanishing axion charge, the bridge disappears and the semiclassical wave function factorizes. In this limit, the theory describes two disconnected expanding branches of de Sitter space with a genuine no-boundary origin, and one can essentially ignore the second copy. This is borne out by the holographic form of the usual Hartle-Hawking no-boundary wave function, which involves a single dual field theory living on a single future boundary \cite{Hertog:2011ky}.

For a finite axion charge, however, the wave function does not factorize. The wormholes prevent the reduction to a single $(d-1)$-sphere and, in striking analogy with the eternal AdS black hole, the two expanding universes would seem to be entangled. Hence a putative dS/CFT description of the no-boundary wave function in this setting would seem to require two duals, one on each future de Sitter boundary, that are somehow entangled. It would certainly be of great interest to study the consequences of this axionic non-factorisation of the Hartle-Hawking wave function in this context in more detail.

Finally, we point out that the axion wormholes that we have considered may also be used to describe axion flux transitions in an existing de Sitter space. In this context, the transition rate involves the difference between the wormhole action and that of the parent background, which yields the expected exponential suppression for larger axion flux. However, a proper discussion of this process, in which axion flux is allowed to vary quantum mechanically, should presumably be based on the interpretation of these solutions as constrained instantons. Another possibly interesting application of these wormholes could be in the search for a derivation of the so-called axionic Festina-Lente Swampland bound \cite{Guidetti:2022xct}, which, it was suggested, might be rooted in the physics of axion-de Sitter wormholes. 

\section*{Acknowledgements}
We thank Jose Barbon, Ben Freivogel, Victor Gorbenko, Edward Morvan, and Martin Sasieta for several discussions, and especially to Simon Maenaut for sharing the code employed for the perturbative stability verification in Appendix \ref{sec:Perturbative Stability}; see \cite{progress} for upcoming work. We benefited from discussions with participants of the “Eurostrings 2023” workshop, where part of this work was presented, as well as the DAMTP workshop "Quantum de Sitter Universe", funded by the Gravity Theory Trust and the Centre for Theoretical Cosmology. SEAG thanks the University of Amsterdam and the Delta Institute for Theoretical Physics for their hospitality and support during the final stages of this project. The work of SEAG, TH, RT, and TVR is partially supported by the KU Leuven C1 grant ZKD1118 C16/16/005, the FWO Research Project G0H9318N and the inter-university project iBOF/21/084. JPVDS is supported by the Delta ITP consortium, a program of the Netherlands Organisation for Scientific Research (NWO) funded by the Dutch Ministry of Education, Culture and Science (OCW).

\appendix

\section{Perturbative stability}\label{sec:Perturbative Stability}
In this Appendix, we verify the perturbative stability of the axion-de Sitter wormholes in four dimensions, by computing the quadratic action of linear fluctuations around the Euclidean background configurations. We follow closely the steps in \cite{Loges:2022nuw} and we will implement the perturbative stability code in \cite{progress} which makes use of the Mathematica packages xPand \cite{Pitrou:2013hga}, xTensor, xPert \cite{Brizuela:2008ra}, and xAct. The main difference with \cite{Loges:2022nuw} is the presence of a cosmological constant term in the action \eqref{eq:onshell fundamental}. 

Working in the Euclidean, the perturbed geometry and axion field read
\begin{align}
    &\rmd s^2 = a(\tau)^2\left[(1+2A)\rmd\tau^2+2\nabla_iB\,\rmd x^i\rmd \tau+[(1-2\psi)\gamma_{ij}+2\nabla_i\nabla_jE]\rmd x^i\rmd x^j\right],\\
    & H_3 = \frac{1}{6}\left(Q+f\right)\varepsilon_{ijk}\rmd x^i\wedge\rmd x^j\wedge\rmd x^k + \frac{1}{2}\varepsilon_{ijk}\nabla^kw\,\rmd\tau\wedge\rmd x^i\wedge\rmd x^j,
\end{align}
where $\gamma_{ij}$ is the metric on $S^3$, $\nabla$ the corresponding covariant derivative and $\varepsilon$ the corresponding Levi-Civita tensor. Also, the closure of $H_3$ entails $\dot{f} = \triangle w$, with $\triangle = \gamma^{ij}\nabla_i\nabla_j$ the Laplacian on $S^3$.

The quadratic action of perturbations involves the field fluctuations, which under a scalar gauge transformation $x^\mu\to x^\mu-\xi^\mu$ with $\xi^\mu=(\xi^\tau,\nabla^i\xi)$ transform as
\begin{align}
    &A\to A + \dot{\xi}^\tau + \mathcal{H}\xi^\tau, && B\to B+\xi^\tau+\dot{\xi}, && \psi\to \psi-\mathcal{H}\xi^\tau\\
    &E\to E+\xi, && f\to f+Q\triangle\xi, && w\to w+Q\dot{\xi}.
\end{align}
Let $\chi=\qty{A,\,B,\,\psi\,,E,\,f,\,w}$ denote any of the field fluctuations. We will now expand them in $S^3$ harmonics as
\begin{equation}
    \chi(\tau,\,x^i)=\sum_{nlm}\chi_{(n)}(\tau)Q^n_{lm}(x^i),
\end{equation}
where $(n)$ denotes $n,\,l,\,m$ collectively, $Q^n_{lm}(x^i)$ are SO$(4)$ spherical harmonics. {For notational simplicity, we will use $1-n^2$ with $n\in\mathbb{N}$ to indicate the eigenvalue of $\triangle$ for the mode $\chi_{(n)}$, and we will suppress the other quantum numbers.}

\begin{figure}[t]
    \centering
    \subfloat[ ]{\includegraphics[width=0.48\textwidth]{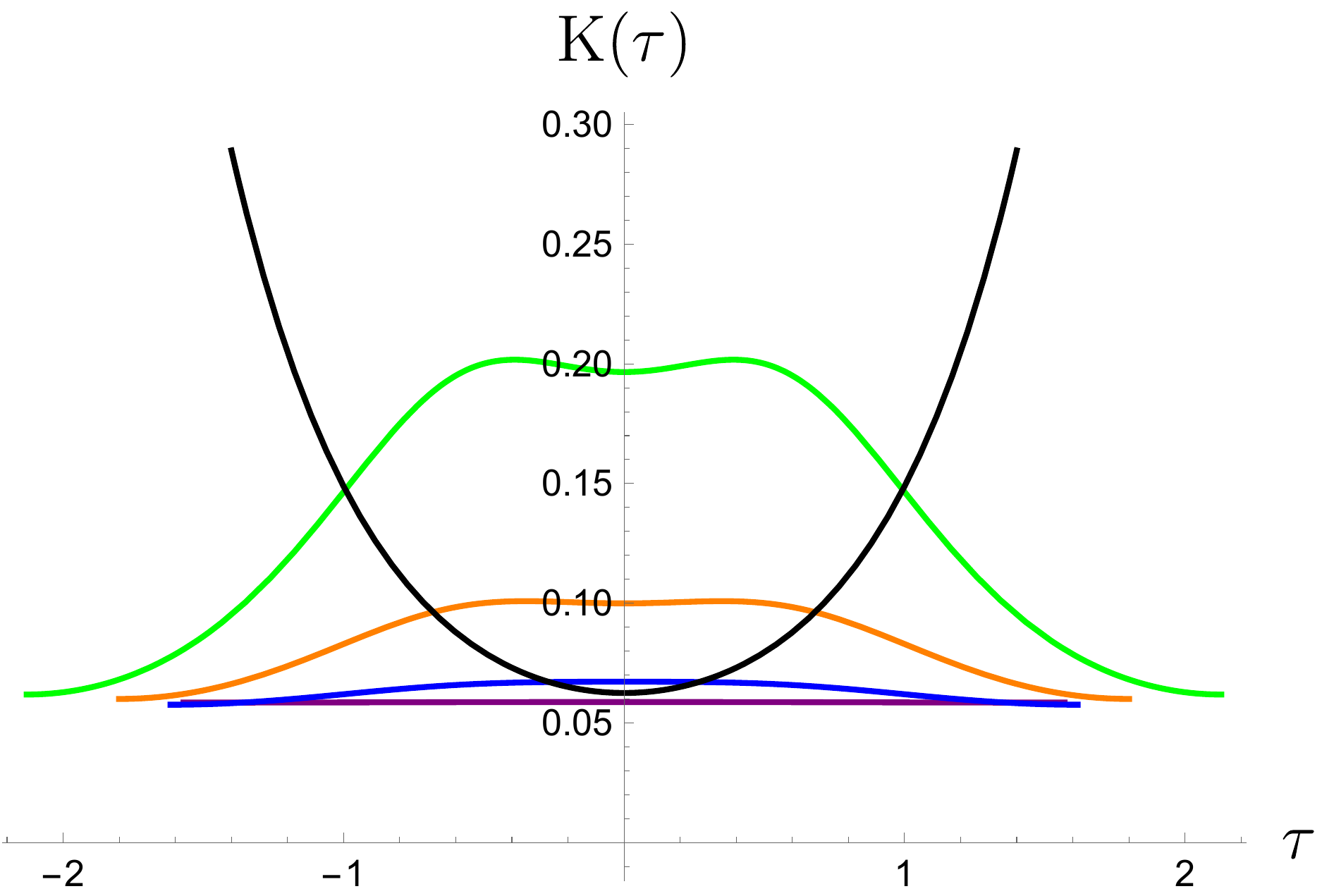}}\hfill \subfloat[]{\includegraphics[width=0.48\textwidth]{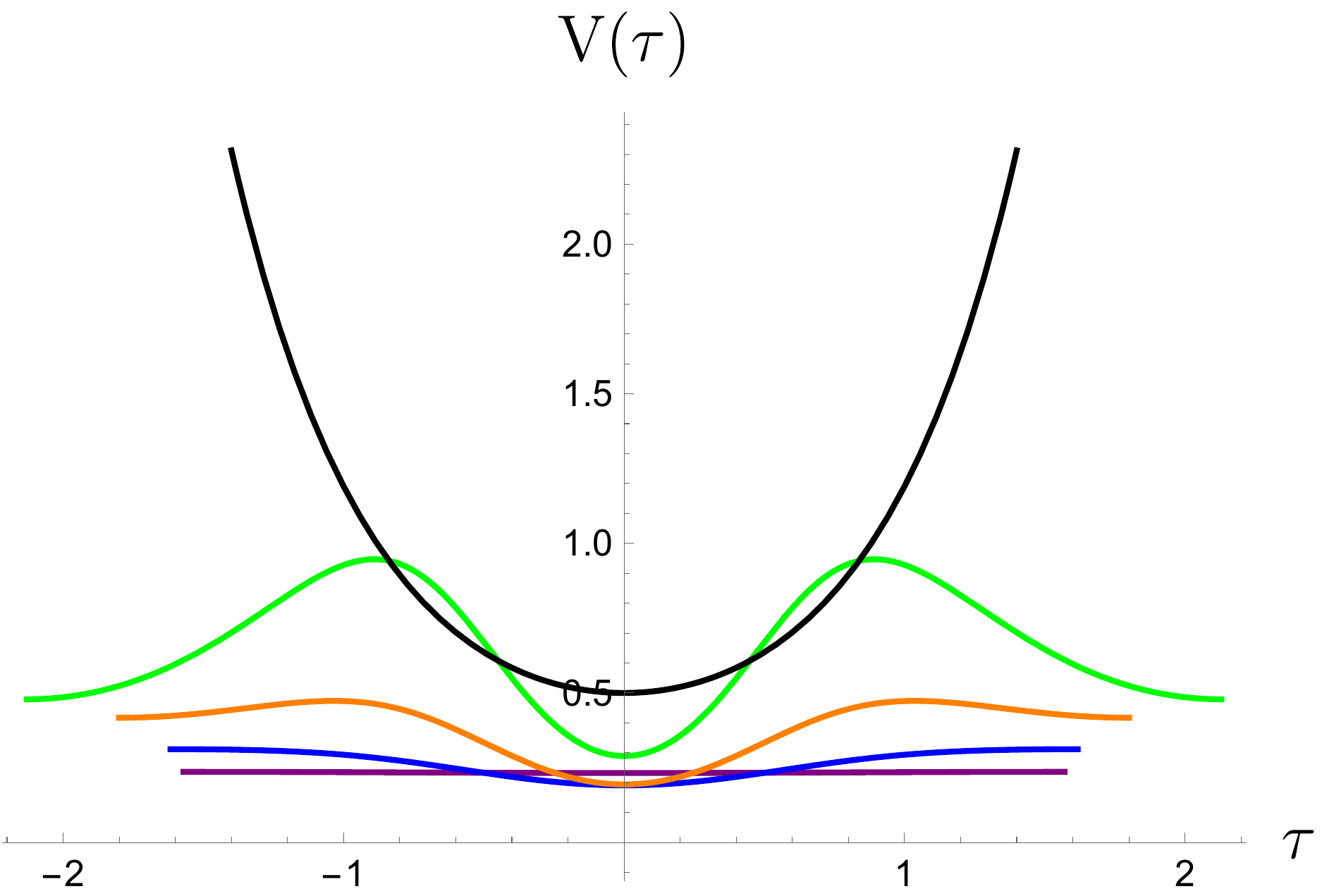}}
    \caption{The behavior of perturbations around axion-de Sitter wormholes is governed by the functions: (a) $K(\tau)$ and (b) $V(\tau)$, shown here for the $n=3$ mode, for different $Q/Q_{\rm max}$ ratios (purple: $1$, blue: $0.9$, orange: $0.6$, green: $0.3$, black: $0$). We see that both functions are everywhere positive, hence the wormholes are perturbatively stable.}
    \label{fig:ceofficients_c}
\end{figure}

Notice also that not all the fluctuations are dynamic degrees of freedom. A subset acts as Lagrange multipliers, the constraints of which are best implemented in the Hamiltonian framework. Following the approach of \cite{Loges:2022nuw}, we introduce the gauge invariant quantity $\mathcal{F}_{(n)}$ and its associated momentum $\Pi_{\mathcal{F}_{(n)}}$ (when $Q\neq Q_{\rm max}$),
\begin{equation}
    \begin{aligned}
    \mathcal{F}_{(n)} &= f_{(n)} - Q(1-n^2) E_{(n)}~,\\ \Pi_{\mathcal{F}_{(n)}} &= \frac{1}{(1-n^2) a^2}\left(Q(1-n^2) B_{(n)} - \dot{f}_{(n)}\right) + \frac{Q}{\mathcal{H} a^2}\psi_{(n)}~.
\end{aligned}
\end{equation}
We also impose the constraints enforced by the Lagrange multipliers $A$ and $B$. The resulting action is manifestly gauge invariant:
\begin{equation}
\delta^2 I_E=\int\rmd\tau\,\qty[\Pi_{\mathcal{F}_{(n)}}\dot{\mathcal{F}}_{(n)}+\mathcal{A}\Pi^2_{\mathcal{F}_{(n)}}-\mathcal{B}\Pi_{\mathcal{F}_{(n)}}\mathcal{F}_{(n)}+\mathcal{C}\mathcal{F}_{(n)}^2],\label{first-order form action}
\end{equation}
where
\begin{equation}\label{eq:coefficients first form}
\begin{aligned}
    \mathcal{A}&=\frac{(1-n^2) a^2}{2}-\frac{3\kappa_4^2Q^2(1-n^2)}{4a^2((1-n^2) +3k)},\\
    \mathcal{B}&=\frac{\kappa_4^2Q^2(1-n^2)}{2a^4((1-n^2) +3k)\mathcal{H}},\\ 
    \mathcal{C}&=\frac{(1-n^2)}{2a^2((1-n^2)+3k)}+\frac{k(3k-\Lambda a^2)}{2a^2\mathcal{H}^2((1-n^2)+3k)},
\end{aligned}
\end{equation}
where $k=1$ is the curvature of the sphere. The momentum $\Pi_{\mathcal{F}_{(n)}}$ appears as a Lagrange multiplier; after integrating $\Pi_{\mathcal{F}_{(n)}}$ out, the action can be written in the second-order form:
\begin{equation}
    \delta^2 I_E=\int\rmd\tau\qty[K\dot{\mathcal{F}}_{(n)}^2+V\mathcal{F}_{(n)}^2]+B\eval{\mathcal{F}_{(n)}^2}_{\tau=\tau_{\text{min}}}^{\tau=\tau_{\text{max}}}~,\label{eq:I_E after momenta integration}
\end{equation}
where we integrated by parts to produce the boundary term. The functions $K,V,$ and $B$ are background-dependent and given by
\begin{align}
    K = -\frac{1}{4\mathcal{A}}, && V = C - \frac{\mathcal{B}^2}{4\mathcal{A}}-\frac{\rmd}{\rmd\tau}\left(\frac{\mathcal{B}}{4\mathcal{A}}\right), && B = \frac{\mathcal{B}}{4\mathcal{A}}.
\end{align}
The periodicity of the boundary conditions means
\begin{equation}
    \mathcal{F}_{(n)}(\tau_{\rm min}) = \mathcal{F}_{(n)}(\tau_{\rm max}).
\end{equation}
Hence the boundary term in \eqref{eq:I_E after momenta integration} vanishes. Since we do not have a closed-form expression for the scale factor $a(\tau)$, we proceed numerically. Figure \ref{fig:ceofficients_c} shows the functions $K (\tau)$ and $V(\tau)$ for the $n = 3$ mode, for a range of different values $Q/Q_{\rm max}$. (The functions $K$ and $V$ for $n>3$ modes are qualitatively the same). It is clear that both functions are always positive across the entire wormhole. Consequently, the quadratic action \eqref{eq:I_E after momenta integration} is non-negative and we can conclude that the wormholes are perturbatively stable.

\section{The arrow of time}\label{sec:Arrow of time}
The universe exhibits a number of arrows of time, but these are all contingent on the fluctuation arrow. The latter is defined by the increase in deviations from homogeneity including, possibly, the formation of cosmic structures \cite{Hartle:2011rb}. A key property of the inflationary cosmologies predicted by the no-boundary wave function is that the fluctuation arrow of time is bidirectional, pointing away from the bounce in both directions \cite{Hartle:2011rb}. The reason is that on the one hand, the no-boundary condition of regularity puts fluctuations in their ground state near the bounce. On the other hand, the inflationary expansion causes perturbations to grow in both directions away from the bounce. In this Appendix, we show that fluctuations around the bouncing axion-de Sitter universes that we have considered behave in a similar fashion.

The growth of the fluctuations $\mathcal{F}_{(n)}$ is governed by the action \eqref{eq:I_E after momenta integration} and its continuation into the Lorentzian regime. We consider separately the cases of generic wormholes with $0<Q<Q_{\rm max}$, and the Nairai limit with $Q=Q_{\rm max}$.

For a slice with vanishing Hubble factor, we perform the Wick rotation $\tau\rightarrow \rmi t$ in the equations of motion resulting from (\ref{eq:I_E after momenta integration}) to obtain the Lorentzian evolution of the axion inhomogeneities:

\begin{equation}\label{eq:Frieddman Lorentz}
    K(t){\mathcal{F}_{(n)}''}(t)+{K}'(t){\mathcal{F}_{(n)}}'(t)+V(t){\mathcal{F}_{(n)}}(t)=0~,
\end{equation}
where the initial value of $\mathcal{F}(t)$ will be fixed by the argument of the wave function on the slice where we make the continuation.

To give intuition about the analytic structure of axion inhomogeneities $\mathcal{F}_{(n)}$, we consider a background where $\kappa_4Q^2/\ell^4\ll 1$. We have confirmed numerically that the solutions with higher-order contributions in the axion charge have a very similar structure. Requiring that the solution to (\ref{eq:Frieddman Lorentz}) is regular at the slice where the continuation is performed, we find
\begin{equation}
        \mathcal{F}_{(n)}(t)=\sec(\frac{t}{\ell})\qty[c^{(1)}_n\sin(\frac{n\, t}{\ell})+c^{(2)}_n\cos(\frac{n\, t}{\ell})]+\mathcal{O}(\kappa^2Q^2/\ell^4)~,
\end{equation}
with $c^{(1)}_n$ and $c^{(2)}_n$ as constants and $t/\ell\leq\tfrac{\pi}{2}$.

\begin{figure}[t]
    \centering
    \includegraphics[width=0.6\textwidth]{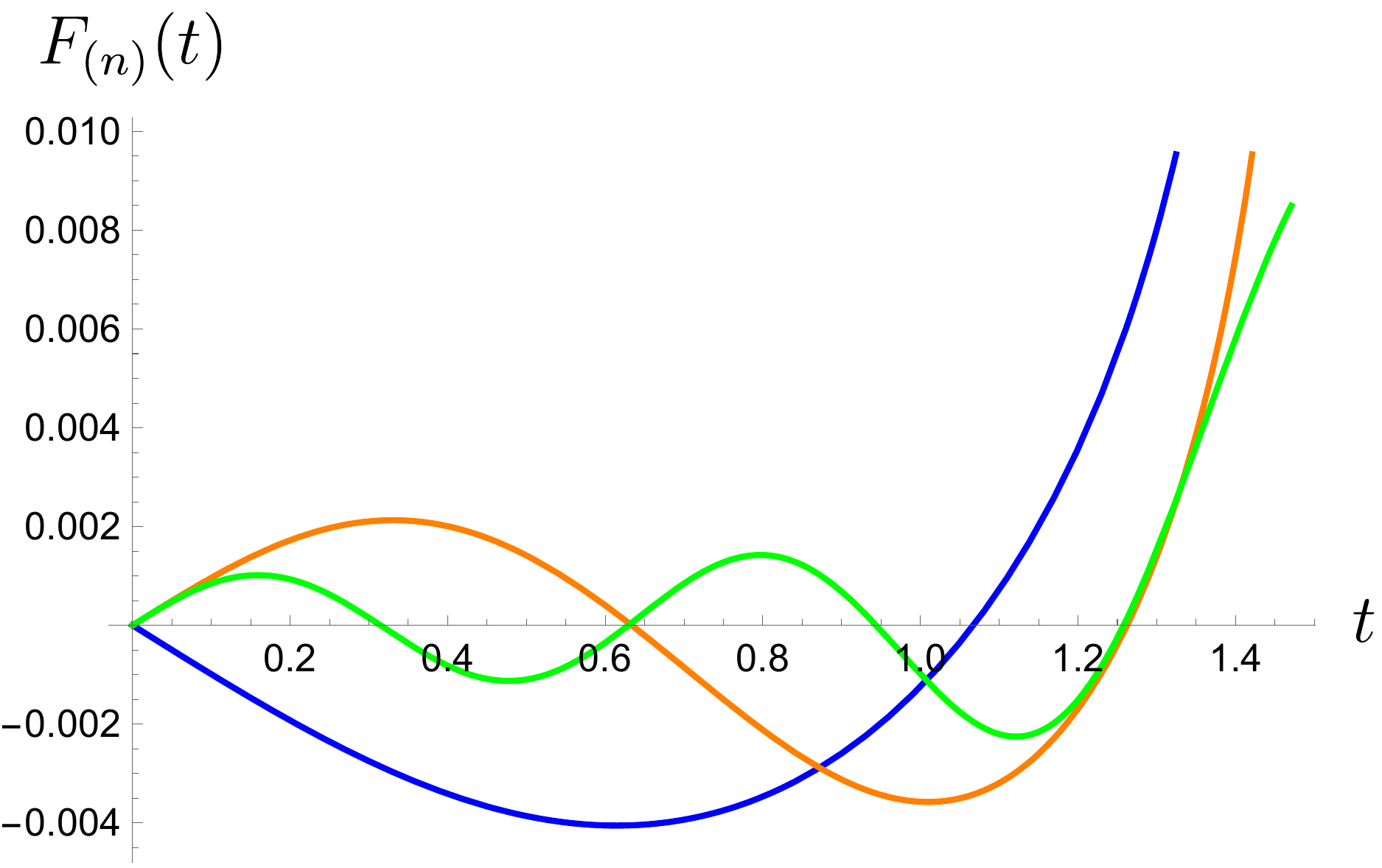}
    \caption{Lorentzian evolution of axionic inhomogeneities for the cosmological constant dominated universe (blue $n=3$, orange $n=5$, green $n=10$).}
    \label{fig:MUfluct}
\end{figure}

In the inflationary context, the fluctuations basically oscillate until the amplitude freezes when the fluctuation mode leaves the horizon. That transition lies at the basis of the arrow of time \cite{Hawking:1993tu}. We find fluctuations around axionic de Sitter wormholes behave in a similar fashion. The fluctuation histories are plotted in \ref{fig:MUfluct}. We see that in all cases, the oscillations cease to exist and instead, the axionic inhomogeneities keep increasing, leading to an emergent fluctuation arrow of time.

In contrast, the Einstein static universe, i.e. $Q=Q_{\rm max}$, shows no arrow of time. The axionic inhomogeneities are always oscillatory given that the scale factor, as well as the coefficients $K$ and $V$ in (\ref{eq:Frieddman Lorentz}) remain constant.

\bibliographystyle{JHEP}
\bibliography{references.bib}
\end{document}